\documentstyle[twoside,psfig]{article}

\catcode`\@=11
\long\def\@makefntext#1{
\protect\noindent \hbox to 3.2pt {\hskip-.9pt  
$^{{\eightrm\@thefnmark}}$\hfil}#1\hfill}		

\def\thefootnote{\fnsymbol{footnote}}
\def\@makefnmark{\hbox to 0pt{$^{\@thefnmark}$\hss}}	
	
\def\ps@myheadings{\let\@mkboth\@gobbletwo
\def\@oddhead{\hbox{}
\rightmark\hfil\eightrm\thepage}   
\def\@oddfoot{}\def\@evenhead{\eightrm\thepage\hfil
\leftmark\hbox{}}\def\@evenfoot{}
\def\sectionmark##1{}\def\subsectionmark##1{}}

\oddsidemargin=\evensidemargin
\renewcommand{\thefootnote}{\fnsymbol{footnote}}

\newcounter{sectionc}\newcounter{subsectionc}\newcounter{subsubsectionc}
\renewcommand{\section}[1] {\vspace{12pt}\addtocounter{sectionc}{1} 
\setcounter{subsectionc}{0}\setcounter{subsubsectionc}{0}\noindent 
	{\tenbf\thesectionc. #1}\par\vspace{5pt}}
\renewcommand{\subsection}[1] {\vspace{12pt}\addtocounter{subsectionc}{1} 
	\setcounter{subsubsectionc}{0}\noindent 
	{\bf\thesectionc.\thesubsectionc. {\kern1pt \bfit #1}}\par\vspace{5pt}}
\renewcommand{\subsubsection}[1] {\vspace{12pt}\addtocounter{subsubsectionc}{1}
	\noindent{\tenrm\thesectionc.\thesubsectionc.\thesubsubsectionc.
	{\kern1pt \tenit #1}}\par\vspace{5pt}}
\newcommand{\nonumsection}[1] {\vspace{12pt}\noindent{\tenbf #1}
	\par\vspace{5pt}}

\newcounter{appendixc}
\newcounter{subappendixc}[appendixc]
\newcounter{subsubappendixc}[subappendixc]
\renewcommand{\thesubappendixc}{\Alph{appendixc}.\arabic{subappendixc}}
\renewcommand{\thesubsubappendixc}
	{\Alph{appendixc}.\arabic{subappendixc}.\arabic{subsubappendixc}}

\renewcommand{\appendix}[1] {\vspace{12pt}
        \refstepcounter{appendixc}
        \setcounter{figure}{0}
        \setcounter{table}{0}
        \setcounter{lemma}{0}
        \setcounter{theorem}{0}
        \setcounter{corollary}{0}
        \setcounter{definition}{0}
        \setcounter{equation}{0}
        \renewcommand{\thefigure}{\Alph{appendixc}.\arabic{figure}}
        \renewcommand{\thetable}{\Alph{appendixc}.\arabic{table}}
        \renewcommand{\theappendixc}{\Alph{appendixc}}
        \renewcommand{\thelemma}{\Alph{appendixc}.\arabic{lemma}}
        \renewcommand{\thetheorem}{\Alph{appendixc}.\arabic{theorem}}
        \renewcommand{\thedefinition}{\Alph{appendixc}.\arabic{definition}}
        \renewcommand{\thecorollary}{\Alph{appendixc}.\arabic{corollary}}
        \renewcommand{\theequation}{\Alph{appendixc}.\arabic{equation}}
        \noindent{\tenbf Appendix \theappendixc #1}\par\vspace{5pt}}
\newcommand{\subappendix}[1] {\vspace{12pt}
        \refstepcounter{subappendixc}
        \noindent{\bf Appendix \thesubappendixc. {\kern1pt \bfit #1}}
	\par\vspace{5pt}}
\newcommand{\subsubappendix}[1] {\vspace{12pt}
        \refstepcounter{subsubappendixc}
        \noindent{\rm Appendix \thesubsubappendixc. {\kern1pt \tenit #1}}
	\par\vspace{5pt}}

\newcommand{\textlineskip}{\baselineskip=13pt}
\newcommand{\smalllineskip}{\baselineskip=10pt}

\def\eightcirc{
\begin{picture}(0,0)
\put(4.4,1.8){\circle{6.5}}
\end{picture}}
\def\eightcopyright{\eightcirc\kern2.7pt\hbox{\eightrm c}} 

\newcommand{\copyrightheading}[1]
	{\vspace*{-2.5cm}\smalllineskip{\flushleft
	{\footnotesize International Journal of Modern Physics D #1}\\
	{\footnotesize $\eightcopyright$\, World Scientific Publishing
	 Company}\\
	 }}

\newcommand{\publisher}[2]{{\begin{center}\footnotesize\smalllineskip 
	Received #1\\
	Revised #2
	\end{center}
	}}

\def\abstracts#1#2#3{{
	\centering{\begin{minipage}{4.5in}\footnotesize\baselineskip=10pt
	\parindent=0pt #1\par 
	\parindent=15pt #2\par
	\parindent=15pt #3
	\end{minipage}}\par}} 

\renewenvironment{thebibliography}[1]
	{\frenchspacing
	 \ninerm\baselineskip=11pt
	 \begin{list}{\arabic{enumi}.}
	{\usecounter{enumi}\setlength{\parsep}{0pt}
	 \setlength{\leftmargin 12.7pt}{\rightmargin 0pt} 
	 \setlength{\itemsep}{0pt} \settowidth
	{\labelwidth}{#1.}\sloppy}}{\end{list}}

\newcounter{itemlistc}
\newcounter{romanlistc}
\newcounter{alphlistc}
\newcounter{arabiclistc}

\newcommand{\fcaption}[1]{
        \refstepcounter{figure}
        \setbox\@tempboxa = \hbox{\footnotesize Fig.~\thefigure. #1}
        \ifdim \wd\@tempboxa > 5in
           {\begin{center}
        \parbox{5in}{\footnotesize\smalllineskip Fig.~\thefigure. #1}
            \end{center}}
        \else
             {\begin{center}
             {\footnotesize Fig.~\thefigure. #1}
              \end{center}}
        \fi}

\newcommand{\tcaption}[1]{
        \refstepcounter{table}
        \setbox\@tempboxa = \hbox{\footnotesize Table~\thetable. #1}
        \ifdim \wd\@tempboxa > 5in
           {\begin{center}
        \parbox{5in}{\footnotesize\smalllineskip Table~\thetable. #1}
            \end{center}}
        \else
             {\begin{center}
             {\footnotesize Table~\thetable. #1}
              \end{center}}
        \fi}

\def\@citex[#1]#2{\if@filesw\immediate\write\@auxout
	{\string\citation{#2}}\fi
\def\@citea{}\@cite{\@for\@citeb:=#2\do
	{\@citea\def\@citea{,}\@ifundefined
	{b@\@citeb}{{\bf ?}\@warning
	{Citation `\@citeb' on page \thepage \space undefined}}
	{\csname b@\@citeb\endcsname}}}{#1}}

\newif\if@cghi
\def\cite{\@cghitrue\@ifnextchar [{\@tempswatrue
	\@citex}{\@tempswafalse\@citex[]}}
\def\citelow{\@cghifalse\@ifnextchar [{\@tempswatrue
	\@citex}{\@tempswafalse\@citex[]}}
\def\@cite#1#2{{$\null^{#1}$\if@tempswa\typeout
	{IJCGA warning: optional citation argument 
	ignored: `#2'} \fi}}

\def\pmb#1{\setbox0=\hbox{#1}
	\kern-.025em\copy0\kern-\wd0
	\kern.05em\copy0\kern-\wd0
	\kern-.025em\raise.0433em\box0}

\def\fnt#1#2{\footnotetext{\kern-.3em
	{$^{\mbox{\scriptsize #1}}$}{#2}}}

\def\thefootnote{\fnsymbol{footnote}}
\def\@makefnmark{\hbox to 0pt{$^{\@thefnmark}$\hss}}	
	
\def\ps@myheadings{%
    \let\@oddfoot\@empty\let\@evenfoot\@empty
    \def\@evenhead{\slshape\leftmark\hfil}
    \def\@oddhead{\hfil{\slshape\rightmark}}
    \let\@mkboth\@gobbletwo
    \let\sectionmark\@gobble
    \let\subsectionmark\@gobble
    }
\font\tenrm=cmr10
\font\tenit=cmti10 
\font\tenbf=cmbx10
\font\bfit=cmbxti10 at 10pt
\font\ninerm=cmr9

\font\eightrm=cmr8

\def\qed{\hbox{${\vcenter{\vbox{			
   \hrule height 0.4pt\hbox{\vrule width 0.4pt height 6pt
   \kern5pt\vrule width 0.4pt}\hrule height 0.4pt}}}$}}

\renewcommand{\thefootnote}{\fnsymbol{footnote}}  

\begin{document}
\setlength{\textheight}{7.7truein}  

\thispagestyle{empty}

\markboth{\protect{\footnotesize\it Special Purpose Pulsar Telescope $\ldots$}}{\protect{\footnotesize\it Special Purpose Pulsar Telescope $\ldots$}}

\normalsize\textlineskip

\setcounter{page}{1}

\copyrightheading{}	

\vspace*{0.88truein}

\centerline{\bf SPECIAL PURPOSE PULSAR TELESCOPE FOR THE}
\vspace*{0.035truein}
\centerline{\bf DETECTION OF COSMIC GRAVITATIONAL WAVES}
\vspace*{0.37truein}

\centerline{\footnotesize Shou-Guan Wang}
\baselineskip=12pt
\centerline{\footnotesize\it National Astronomical Observatories,
                Chinese Academy of Sciences,}
\baselineskip=10pt
\centerline{\footnotesize\it Beijing 100012, China}
\vspace*{10pt}
\centerline{\footnotesize Zong-Hong Zhu\footnote{Email: zong-hong.zhu@nao.ac.jp~~~~~~zhuzh@bao.ac.cn}}
\baselineskip=12pt
\centerline{\footnotesize\it National Astronomical Observatories,
                Chinese Academy of Sciences,}
\baselineskip=10pt
\centerline{\footnotesize\it Beijing 100012, China}
\centerline{\footnotesize\it National Astronomical Observatory,}
\baselineskip=10pt
\centerline{\footnotesize\it 2-21-1, Osawa, Mitaka, Tokyo 181-8588, Japan}
\vspace*{10pt}
\centerline{\footnotesize Zhen-Long Zou}
\baselineskip=12pt
\centerline{\footnotesize\it National Astronomical Observatories,
                Chinese Academy of Sciences,}
\baselineskip=10pt
\centerline{\footnotesize\it Beijing 100012, China}
\vspace*{10pt}
\centerline{\footnotesize Yuan-Zhong Zhang}
\baselineskip=12pt
\centerline{\footnotesize\it Institute of Theoretical Physics,    
                Chinese Academy of Sciences,}
\baselineskip=10pt
\centerline{\footnotesize\it Beijing 100080, China}
\vspace*{0.225truein}
\publisher{(received date)}{(revised date)}

\vspace*{0.21truein}
\abstracts{ 
Pulsars can be used to search for stochastic backgrounds of gravitational waves
of cosmological origin within the very low frequency band (VLF), $10^{-7}$ to
$10^{-9}$ Hz.  We propose to construct a special 50 m radio telescope.
Regular timing measurements of about 10 strong millisecond 
pulsars will perhaps allow the detection of gravitational waves within VLF
or at least will give a more stringent upper limits. 
}{}{}


\vspace*{1pt}\textlineskip      
\section{Introduction}          
\vspace*{-0.5pt}
\noindent
Because a background of relic particles gives a snapshot of the state of the
universe at the time when these particles decoupled from the primordial plasma,
relic gravitational waves (gravitons) are a potential source of informations
on the state of the very early universe and the physics at correspondingly
high energies, which can not be accessed experimentally in any other 
way\cite{mag00}.
The photons of the cosmic microwave background (CMB) give us a snapshot of the 
state of the universe at $t \sim 3\times 10^5$ years after the big-bang, while 
the gravitons of the stochastic gravitational wave background (SGWB) encode in 
its spectrum the information of the universe at $t \sim 10^{-44}$ seconds.
These properties make the detection of SGWB very interesting, which is the 
main purpose of the proposal discussed here.

A SGWB can be characterized by its energy distribution in frequency, especially
the dimensionless quantity
\begin{equation}
\Omega_{\rm gw}(f):={1\over\rho_{\rm critical}}\
{d\rho_{\rm gw}\over d\ln f}\ ,
\end{equation}
where $\rho_{\rm critical}=3c^2H_0^2/8\pi G$ is the critical energy density 
required (today) to close the universe 
(the Hubble constant 
$H_0=100h_{100}$ km sec$^{-1}$ Mpc$^{-1}$ = $3.2\times 10^{-18} h_{100}$ Hz 
). 
Up to now, there have been three observational constraints on SGWB 
spectrum\cite{all96,all99}:
(i) The strongest observational constraint on $\Omega_{\rm gw}(f)$ comes
from the high degree of isotropy observed in the CMB,
$
\Omega_{\rm gw}(f)\  h_{100}^2 < 7 \times 10^{-11} \left({H_0/f} \right)^2
\quad{\rm for}\quad H_0 < f < 30 H_0\ .
$
(ii) The second comes from the standard model of big-bang nucleosynthesis,
$
\int_{f>10^{-8}\ {\rm Hz}} d\ln f\ \Omega_{\rm gw }(f)\ h_{100}^2 < 10^{-5}\ .
$
(iii) The last observational constraint comes from monitoring the radio pulses 
emitted by a number of stable millisecond pulsars for a decade,
$
\Omega_{\rm gw}(f=10^{-8}\ {\rm Hz})\ h_{100}^2 < 10^{-8}\ .
$

In order to get a lower constraint on SGWB spectrum or even
directly detect SGWB within the very low frequency band (VLF), $10^{-7}$ to
$10^{-9}$ Hz, we propose to construct a special 50 m radio telescope at National
Astronomical Observatories, Chinese Academy of Sciences (NAOC) to monitor about 
10 strong millisecond pulsars. 
After reviewing the basics of pulsars as probes for gravitational wave 
(section 2), we present our research plan in details (section 3).

\setcounter{footnote}{0}
\renewcommand{\thefootnote}{\alph{footnote}}

\section{Pulsar timing array as efficient gravitational wave detectors}
\noindent
Pulsars have been believed to be extremely stable natural clocks.
Detweiler\cite{det79} had shown that measurements of pulse arrival time (TOA) 
may be used to search for gravitational waves with a
period on the order of years. For SGWB, his found  
\begin{equation}
\rho_{\rm gw} = \frac{243 \, \pi^3 \, f^4 \, \langle R^2(t) \rangle}{208 \, G}.
\end{equation}
where the residual $R(t)$ is the difference between observed and expected 
TOA. The span of measurements T should be longer than $1/f$ 
(with $1/f =$ a few years, the measurements may be taken at intervals of say, 
10-20 days).

The observed $R(t)_{\rm obs}$ consists of deviations caused by various effects.
Some of them are systematic, such as the uncertainties in the position and 
proper motion of pulsars, errors in the positions of barycenter. 
While others are random, for example, propagation delays caused by interstellar 
turbulence, relativistic effects by intervening massive objects, irregularities 
occurred in pulsar parameters, or in reference clock. 
Some of these effects could be modeled or be estimated, and hence be corrected. 
The resulted $\langle R^2(t) \rangle$  is therefore the 
mixture of three components: (1) the remaining effects left over after the 
correction; (2) pulse timing noise which is related with (a) signal-to-noise 
ratio of pulsar observation, (b) pulse shape and the definition of arrival 
time, and (c) method in obtaining the arrival time; (3) background 
gravitational waves. 
Thus the quantity $\langle R^2(t) \rangle$ from pulsar timing observations
produces an upper limit on the energy density of SGWB.

Following Detweiler's work, a number of observations had been made over a 
decade due to the advent of the discovery of millisecond pulsars.
Millisecond pulsars have high stability and sharp pulses, therefore 
the precision of timing measurements is extraordinary high, reaching 
$0.1\,\mu{\rm s}$ sometimes. 
A millisecond pulsar, besides its extraordinary high inertia which warrants 
high stability, the shape of its pulses being so sharp has caused a great 
increase in the precision of timing measurements. 
As a result, the energy density of SGWB in the frequency range around 
0.5 cycle/year was found to be $\rho < 10^{-6}\rho_{\rm critical}$ 
(see eg. Ref~5).

This result is encouraging. 
In particular, a much better sensitivity will be obtained if one can use two
or more pulsars as coincident detectors\cite{hel83,fos90,man93,lor01}.
Such a ``pulsar timing array'' could be a very efficient SGWB detector.
To quantify this, consider the Doppler shift from the $i$th pulsar in the 
array\cite{lor01}
\begin{equation}
\frac{\Delta\nu_i (t)}{\nu_i}=\alpha_i s(t)+n_i(t)
\end{equation}
where $s(t)$ is the GW signal, $\alpha_i$ a geometrical factor
depending on the line-of-sight direction, and $n_i(t)$ contains the
noise intrinsic to the pulsar timing. The cross-correlation between
two pulsars then gives\cite{mag00}
\begin{equation}
\alpha_i\alpha_j\langle s^2(t)\rangle +\alpha_i\langle sn_i\rangle
+\alpha_j\langle sn_j\rangle +\langle n_in_j\rangle
\end{equation}
and, since the timing noises from different pulsars are uncorrelated,
increasing the observation time will limit the cross-correlation towards
$\alpha_i\alpha_j\langle s^2(t)\rangle$. 

Unfortunately, the galactic millisecond pulsars known around 1990, for which 
timing data have been taken already, were concentrated in almost the same
region of the sky, and correlation analysis of these data does not improve 
much the result.
Now the known millisecond pulsars are much more uniformly
distributed on the sky.
It is the right time to construct such a ``pulsar timing array''\cite{lor01}.
The Berkeley pulsar group led by Don Backer has been working on this 
direction for a few years.

\section{A special 50 m radio telescope at NAOC}
\noindent
We propose to construct a special 50m radio telescope at NAOC, 
devoted to pulsar timing measurement. 
It will work on low frequencies, with 1400 MHz as the highest (so that 
the instruments will not be very expensive). 
At least 10 strong millisecond pulsars can be chosen from the 27 candidates 
listed in Table~1 for regular timing measurements. 
Each object should be observed 1 hour every day. 
With the best receiving system today, accumulation of 15 day's measurements 
will have a sensitivity of about 0.05 mJy, so that a signal-to-noise ratio 
above 10 is expected for these pulsars.    

\eject
\begin{table}[htbp]
{
\centering
\tcaption{\sf 27 candidates to be observed with 50 m telescope (S400 $>$ 10mJy)}
\centerline{\footnotesize\smalllineskip
\begin{tabular}{l r r l l l}\\
\hline
2000Name   &  Gal-L & Gal-B  &  Period (ms)      & Distance(kpc) &  S400(mJy)\\
\hline
J0034-0534 & 111.49 & -68.07 &  1.8771818543796  & 0.98          &  17       \\
J0613-0200 & 210.41 & -9.30  &  3.06184403674401 & 2.19          &  2e+01    \\
J0751+1807*& 202.73 & +21.09 &  3.478770781510   & 2.02          &  1e+01    \\
J1012+5307 & 160.35 & +50.86 &  5.25574901198    & 0.517         &  3e+01    \\
J1022+1001*& 231.79 & +51.10 & 16.452929681440   & 0.599         &  2e+01    \\
J1300+1240*& 311.30 & +75.41 &  6.2185319388187  & 0.624         &  2e+01    \\
J1455-3330 & 330.72 & +22.56 &  7.987204795504   & 0.738         &   9.0     \\
J1623-2631 & 350.98 & +15.96 & 11.075750876440   & 1.8           &  2e+01    \\
J1643-1224 &   5.67 & +21.22 &  4.62164144656360 & 4.86          &  8e+01    \\
J1713+0747*&  28.75 & +25.22 &  4.57013652273380 & 0.892         &  4e+01    \\
J1730-2304 &   3.14 &  +6.02 &  8.1227979128499  & 0.506         &  4e+01    \\
J1744-1134 &  14.79 &  +9.18 &  4.07454587512695 & 0.166         &  18       \\
J1748-2446A&   3.84 &  +1.70 & 11.563148389585999401&7.1         &  -        \\
J1804-2717 &   3.51 &  -2.74 &  9.343030681150   & 1.17          &  15       \\
J1823-3021A&   2.79 &  -7.91 &  5.440002278840   & 8             &  16       \\
J1824-2452 &   7.80 &  -5.58 &  3.0543146293258  & 5.5           &  4e+01    \\
J1857+0943*&  42.29 &  +3.06 &  5.36210045404154 & 1             &  3e+01    \\
J1911-1114 &  25.14 &  -9.58 &  3.6257455713977  & 1.59          &  31       \\
J1939+2134*&  57.51 &  -0.29 &  1.5578064924327  & 9.65          &  2e+02    \\
J1955+2908 &  65.84 &  +0.44 &  6.133166488729   & 5.39          &  2e+01    \\
J1959+2048 &  59.20 &  -4.70 &  1.60740168480632 & 1.53          &  2e+01    \\
J2019+2425*&  64.75 &  -6.62 &  3.9345240796636  & 0.912         &  2e+01    \\
J2051-0827 &  39.19 & -30.41 &  4.50864174335    & 1.28          &  2e+01    \\
J2124-3358 &  10.93 & -45.44 &  4.93111485914810 & 0.248         &  17       \\
J2145-0750 &  47.78 & -42.08 & 16.0524236584091  & 0.5           &  100      \\
J2229+2643*&  87.69 & -26.28 &  2.9778192947192  & 1.43          &  1e+01    \\
J2317+1439*&  91.36 & -42.36 &  3.4452510710225  & 1.89          &  2e+01    \\
\hline\\
\end{tabular}}
}
\footnotesize {*--}{Being observed by USA groups}
\end{table}

In addition to what mentioned above, multi-object observation strategy 
is of importance 
for the detection of SGWB in the following senses: Consider a number of pulsars 
separated at different parts in the sky, gravitational wave from any direction 
sweeping across the Earth will be marked by the change of the rates of local 
clock at a frequency $f$. Owing to the quardrupole character of the 
gravitational radiation, such effect will exactly in two pulsar data if 
their directions are $180^\circ$ apart. 
So that $R(t)$, in the form of random fluctuations, contained in the record of 
timing measurements of a pulsar would 
have an exactly similar shape with the $R(t)$ record of a pulsar. 
As the period of the gravitational wave $1/f$ being very long, observations of 
two pulsars within a time interval of one day can be thought as nearly 
simultaneous. $R(t)_{\rm obs}$ of the two records, when cross-correlated, 
will reveal the existence quantitatively of the background gravitational waves. In practice, we will have a number of pairs of pulsars at different 
separation angles, by the same principle, cross-correlations performed to the 
$R(t)_{\rm obs}$ of all the pairs will produce a better result than that from
a single pair.

Finally, we would like to remark that, among these sources of perturbations 
contained in $R(t)$, many of them are of interest in their own right. 
The timing measurement data of these objects may be used to, for examples,
(1)improve the ephemeris, (2)monitor the intrinsic perturbations of pulsars,
(3)establish a time standard for long time scales.

\nonumsection{Acknowledgements}
\noindent
The authors are grateful to Prof. JinLin Han for providing us the data in 
Table~1 and careful reading and improving the manuscript,
and also to Prof. Wei-Tou Ni for helpful comments and suggestions.
Z.H.Zhu is supported
  by the Special Funds for Major State Basic Research Projects, 
  the National Natural Science Foundation of China (19903002) 
  and the JSPS fellowship of Japan.

\nonumsection{References}

\end{document}